\documentstyle[epsfig,newpasp,twoside]{article}
\def\etal{{\em et al.~}}
\def\be{\begin{equation}}
\def\ee{\end{equation}}
\def\bea{\begin{eqnarray}}
\def\eea{\end{eqnarray}}

\def\etal{{\em et al.~}}

\def\mum{\mu {\rm m}}
\def\kms{{\rm \,km\,s}^{-1}}

\def\pster{{\rm \,ster}^{-1}}

\def\dg{^{\circ}}

\def\Jy{{\rm \,Jy}}
\def\MJy{{\rm \,MJy}}
\def\spose#1{\hbox to 0pt{#1\hss}}
\def\simlt{\mathrel{\spose{\lower 3pt\hbox{$\mathchar"218$}}
     \raise 2.0pt\hbox{$\mathchar"13C$}}}
\def\simgt{\mathrel{\spose{\lower 3pt\hbox{$\mathchar"218$}}
     \raise 2.0pt\hbox{$\mathchar"13E$}}}
\def\({\left(}
\def\){\right)}
\def\[{\left[}
\def\]{\right]}
\def\<{\left\langle}
\def\>{\right\rangle}

\def\IAU130{in {\em Large Scale Structures of the Universe}, IAU Symposium 130~}
\def\MN{{\em Mon.Not.R.astr.Soc.~}}

\def\edcomment#1{\iffalse\marginpar{\raggedright\sl#1\/}\else\relax\fi}
\reversemarginpar

\begin{document}
\title{The Behind The Plane Survey - source selection, identifications and completeness}
\author{Will Saunders$^\ast$, Kenton D'Mellow$^\ast$, Brent Tully$^\dag$, Esperanza Carrasco$^\dag$}
\affil{$^\ast$University of Edinburgh, UK. $^\dag$University of Hawaii, USA. $^\ddag$INAOE, Mexico.}
\author{Bahram Mobasher$^\ast$, Steve Maddox$^\dag$, George Hau$^\ddag$}
\affil{$^\ast$STSCI, USA. $^\dag$University of Nottingham, UK. $^\ddag$Universidad Catholica, Santiago, Chile.}
\author{Will Sutherland$^\ast$, Dave Clements$^\dag$, Lister Staveley-Smith$^\ddag$}
\affil{$^\ast$ATC, ROE, UK. $^\dag$University of Cardiff, UK. $^\ddag$CSIRO, Australia.}

\begin{abstract}

We present details of the Behind The Plane survey of IRAS galaxies,
which extends the PSCz survey to cover the 93\% of the sky with
complete and reliable IRAS data from the Point Source Catalog. At low
latitudes, our catalogue is not complete to $0.6\Jy$, but the
incompleteness is physically understood and can be corrected for. IRAS galaxies at low latitudes are heavily or
completely obscured optically, and are heavily outnumbered by Galactic
sources with similar IRAS properties. We have used radio, optical, mm,
near and far-infrared data to identify the galaxies, and the 2D catalogue is now complete. We have used optical and
HI spectroscopy to obtain redshifts for the galaxies; the southern spectroscopy is completed and the north nearly so.

\end{abstract}
\vspace{-10pt}

\section{Introduction}
\vspace{-5pt}
The 84\% sky coverage of the PSCz survey is effectively limited by the need to get, for every galaxy, an optical identification from sky survey plates. The IRAS Point Source Catalog data itself is reliable to much lower latitudes, although genuine galaxies are outnumbered by Galactic sources with similar IRAS properties. Previous attempts to go further into the Plane have either been restricted to the Arecibo declination range, or have relied on optical identifications from Sky Survey Plates. Because the extinction may be several magnitudes, they have inevitably suffered from progressive and unquantifiable incompleteness as a function of latitude. In 1994 we embarked on a program, parallel with the PSCz survey, to systematically identify low latitude IRAS galaxies wherever the PSC data allowed, using new near-infrared observations where necessary. 

\vspace{-5pt}
\section{Sky coverage and selection criteria}
\vspace{-5pt}
The mask consists of (a) the IRAS coverage gaps (3\% of the sky), (b) areas flagged as High Source Density at $60\mum$ (3\%), where the PSC processing was changed to ensure reliability at the expense of completeness, (c) areas flagged as High Source Density at $12$ or $25\mum$ on the basis that identifications would be impossible, and (d) areas with $I_{100} > 100 \MJy\pster$ (as in Rowan-Robinson \etal 1991), because of overwhelming contamination by local sources. Our final sky coverage is 93\%.

The IRAS selection criteria were tightened from those used for the PSCz, in order to minimise the contamination by Galactic sources while still keeping most of the galaxies. Upper limits were only used where they guaranteed inclusion or exclusion. The revised criteria were \\

\begin{tabular}{lcc} 
$S_{60}/S_{25}$ &$>$& $2$ \\
$S_{60}/S_{12}$ &$>$& $4$ \\
$S_{100}/S_{60}$ &$>$& $1$ \\
$S_{100}/S_{60}$ &$<$& $5$ \\
$CC_{60}$ &$>$ & 97.5\%
\end{tabular}

\vspace{-5pt}
\section{Identifications}
\vspace{-5pt}
Many sources are immediately identifiable as galaxies from sky survey plates. Others are clearly Galactic. To identify the rest, we used sky survey plates in all available bands, VLA NVSS data for $\delta>-40\dg$, existing submm observations, IRAS addscan profiles, Simbad and other literature data. For almost all sources still remaining unclassified, and also almost all sources with a faint ($r>18.5^m$) galaxy counterpart, we obtained $K'$ `snapshots' using the UHa $88''$, UNAM 2.1m, ESO 2.2m, CTIO 1.5m and Las Campanas 1m telescopes, between 1994 and 1999. We also included in this imaging program any low-latitude PSCz sources whose identification was unclear, since the PSCz contains known incompleteness at low latitudes (Saunders \etal 2000).

In general, the $K'$ images allowed unambiguous identification as a galaxy or Galactic source. Occasionally, there remained ambiguity between galaxies and buried YSO's, and more frequently a very faint and compact galaxy ID had been overlooked, but revealed by subsequent NVSS data. Most galaxies have some sort of optical counterpart visible on sky survey plates, though several hundred do not. Our identification program gave us a total 1225 galaxies. 

\vspace{-5pt}
\section{Completeness of the BTP survey}
\vspace{-5pt}
There is incompleteness in the BTP at all fluxes due to our colour criteria, which are more restrictive than for the PSCz and inevitably exclude some galaxies with unusual colours. There is also incompleteness because the BTP relies on point source filtered fluxes: nearby galaxies have their flux underestimated and may drop out of the sample. This is actually rather rare, since these nearby galaxies typically have fluxes well above the flux limit. These effects can combine: the different beam sizes at different wavebands can give nearby galaxies strange colours. Based on the colours and fluxes of PSCz galaxies, we estimate these combined incompletenesses at 10\% with weak dependence on flux or distance. Sadly, this incompleteness includes Dwingeloo I (Kraan-Korteweg \etal 1994) and possibly other very nearby galaxies.

The nominal flux limit for the BTP survey is $0.6\Jy$, as for the PSCz. However, the source counts show obvious incompleteness to this limit. The principal cause of this is simply low signal-to-noise, and in particular the requirement for each individual source detection to be $3\sigma$ (Beichman \etal 1988, henceforth ES, V.C.III). In areas scanned just after crossing the Plane, this was exacerbated by the lagging noise estimator and consequent source shadowing (ES V.C.2, VIII.D.6). 

The PSC quotes a s/n for each source, this being the lowest s/n on any detection. Figure 1 shows the source counts for the PSCz and BTP surveys as a function of the quoted s/n, categorised by the noise level itself, for all sources with a quoted noise level greater than $0.2\Jy$. The source counts show very similar behaviour as a function of signal-to-noise for all noise levels. We have developed a simple model for this incompleteness, motivated by the requirement that $CC_{60} > 97.5\%$ and hence that the s/n on the {\em best} detection be at least 6.2 (ES V.C.4). For an average of 6 detections/source, we then expect the range in s/n amongst the detections to approximately follow a Gaussian distribution with mean 2.6 and scatter 0.82 (e.g. Kendall \& Stuart 1973). In this case the incompleteness should be well described by an error function and this appears to be a reasonable description for the data (Figure 1), at least above s/n=3.35. We assume that sources are undetectable below s/n=3.35, and assign s/n=3.4 to the handful of sources actually detected below this level. This is the only adjustable parameter in our model. We thus have, for each source, a weight equal to the inverse of the determined incompleteness for that source, and a flux limit corresponding the the s/n limit; together these allow us to determine the selection function for each source.

\begin{figure}
\centerline{\epsfig{figure=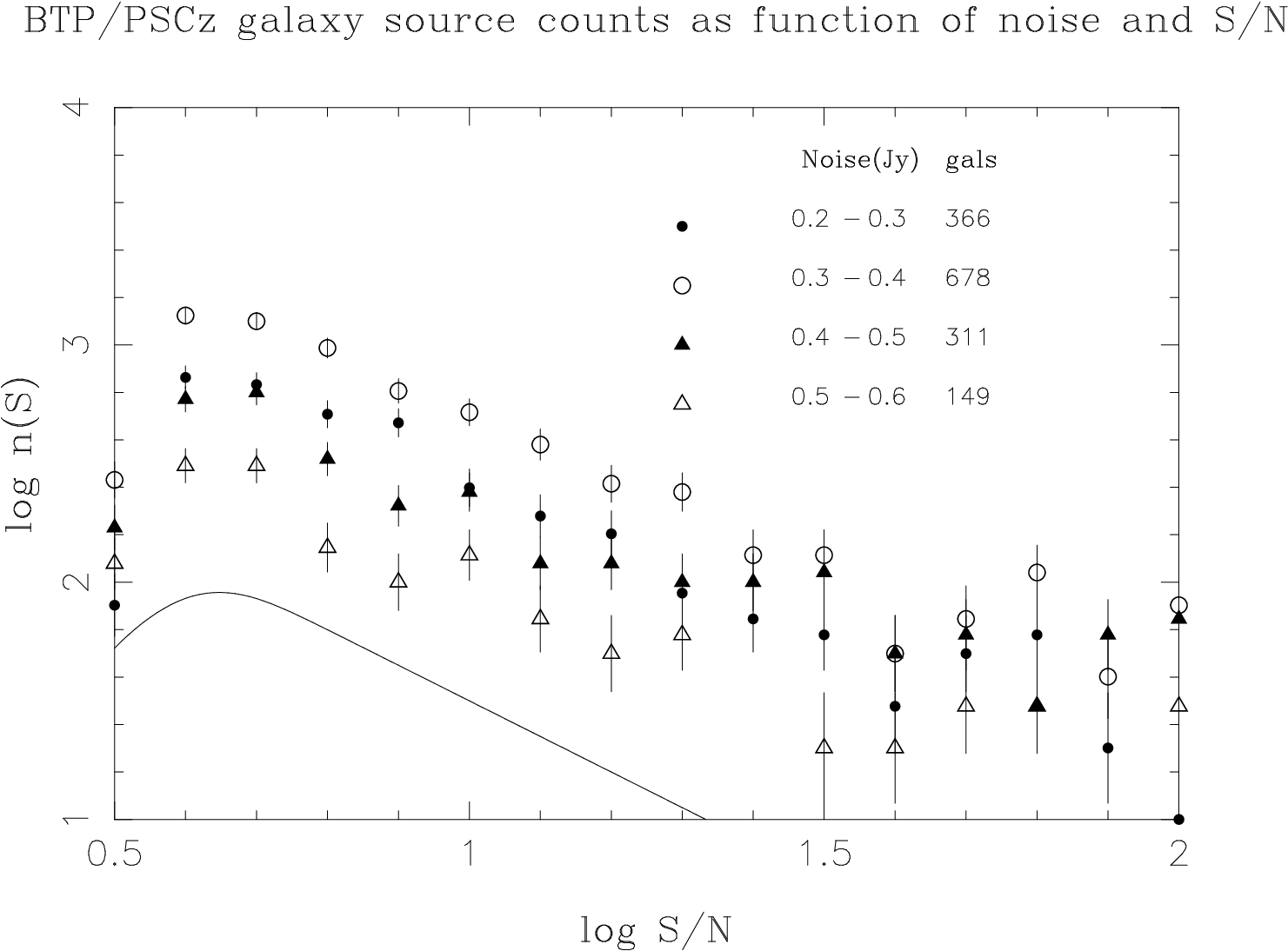,width=9cm,angle=-90}}
Figure 1. PSCz and BTP galaxy source counts as a function of their signal-to-noise, for different noise levels. Vertical normalisation is arbitrary. The line shows Euclidean counts convolved with the incompleteness function described in the text.
\end{figure}

\vspace{-5pt}
\section{Redshift aquisition}
\vspace{-5pt}
About 40\% of the required redshifts are known from other surveys. Most sources with a galaxy ID from the K' imaging have a counterpart on sky survey plates, though often only just. Galaxies fall below the plate limit either because their surface brightness is too low or because their total magnitude is too faint.  The former are typically nearby, normal spiral galaxies, accessible to 21cm spectroscopy. The latter are usually very distant, FIR-luminous galaxies, barely resolved from stars in our K'-images and with estimated distances beyond our redshift completeness target of $25,000 \kms$. The southern spectroscopy for both BTP and low-latitude PSCz is now complete, using Parkes and the CTIO 1.5m and 4m telescopes, though some data is unreduced. In the north, we have obtained 250 redshifts at the INAOE 2.1m; we have about 150 redshifts still to obtain, and we hope to complete the programme this autumn.

\end{document}